\documentclass[a4paper,conference]{IEEEtran}
\usepackage{graphicx}
\usepackage{caption}
\usepackage{cite}
\usepackage{caption}
\usepackage{amssymb}
\usepackage{hyperref}

\ifCLASSINFOpdf
\else
\fi
\begin{document}
%
% paper title
% Titles are generally capitalized except for words such as a, an, and, as,
% at, but, by, for, in, nor, of, on, or, the, to and up, which are usually
% not capitalized unless they are the first or last word of the title.
% Linebreaks \\ can be used within to get better formatting as desired.
% Do not put math or special symbols in the title.
\title{Spatio-Temporal Analysis of Transformer based Architecture for Attention Estimation from EEG}

% author names and affiliations
% use a multiple column layout for up to three different
% affiliations

\author{
    \IEEEauthorblockN{
        Victor Delvigne \IEEEauthorrefmark{1} \IEEEauthorrefmark{2},
        Hazem Wannous \IEEEauthorrefmark{2}, 
        Jean-Philippe Vandeborre \IEEEauthorrefmark{2},
        Laurence Ris \IEEEauthorrefmark{3}, 
        Thierry Dutoit \IEEEauthorrefmark{1}
    }
    \IEEEauthorblockA{
        \IEEEauthorrefmark{1} ISIA Lab, Faculty of Engineering, \textit{University of Mons}, Belgium; \\
        \IEEEauthorrefmark{2} IMT Nord Europe, \textit{CRIStAL UMR CNRS 9189}, France; \\
        \IEEEauthorrefmark{3} Neuroscience Lab, Faculty of Medecine and Pharmacy, \textit{University of Mons}, Belgium
    }
    Email: victor.delvigne@umons.ac.be
}

\maketitle

\begin{abstract}
For many years now, understanding the brain mechanism has been a great research subject in many different fields. Brain signal processing and especially electroencephalogram (EEG) has recently known a growing interest both in academia and industry. One of the main examples is the increasing number of Brain-Computer Interfaces (BCI) aiming to link brains and computers. In this paper, we present a novel framework allowing us to retrieve the attention state, i.e degree of attention given to a specific task, from EEG signals. While previous methods often consider the spatial relationship in EEG through electrodes and process them in recurrent or convolutional based architecture, we propose here to also exploit the spatial and temporal information with a transformer-based network that has already shown its supremacy in many machine-learning (ML) related studies, e.g. machine translation. In addition to this novel architecture, an extensive study on the feature extraction methods, frequential bands and temporal windows length has also been carried out. The proposed network has been trained and validated on two public datasets and achieves higher results compared to state-of-the-art models. As well as proposing better results, the framework could be used in real applications, e.g. Attention Deficit Hyperactivity Disorder (ADHD) symptoms or vigilance during a driving assessment.
\end{abstract}

% no keywords

% For peer review papers, you can put extra information on the cover
% page as needed:
% \ifCLASSOPTIONpeerreview
% \begin{center} \bfseries EDICS Category: 3-BBND \end{center}
% \fi
%
% For peerreview papers, this IEEEtran command inserts a page break and
% creates the second title. It will be ignored for other modes.
\IEEEpeerreviewmaketitle

\section{Introduction}
Nowadays, deep learning (DL) and other ML algorithms have known a huge increase in interest that has led to improvements in several scientific fields. Different domains have benefited from ML researches such as natural language processing (NLP), computer vision, speech recognition or understanding. However, another field where the use of DL remains elusive is brain imaging, the goal of these models being to help to better understand the mechanism within the brain. The considered signals can represent the brain matter composition with magnetic resonance imaging (MRI) \cite{lundervold_overview_2019} or the electrical activity of its neurons with EEG \cite{tirupattur_thoughtviz_2018}. The goal of the model considering EEG are to evaluate human cognitive faculties to have a better understanding of brain function.  

Recent works propose to consider electrophysiological signals and especially EEG to estimate the attentional state of participant \cite{delvigne_phydaa_2021,cao_multi-channel_2019,zheng_multimodal_2017,gao_eeg-based_2019}, i.e. a metric expressing the ability of an individual to be concentrate on a given task. The purposes of these researches are wide and could help in various fields: medical, entertainment, road safety or marketing. The proposed approaches consider the use of ML models that have already shown outperforming results in other fields: fully-connected neural networks \cite{zou_eeg-based_2020}, convolutional neural networks (CNN) \cite{gao_eeg-based_2019}, recurrent neural networks (RNN) \cite{delvigne_phydaa_2021} or combinations of these last. Moreover, the models can consider extracted features from EEG \cite{zheng_multimodal_2017} or preprocessed EEG directly \cite{gao_eeg-based_2019}.

Although these works present promising results, they tend to ignore some of the sequential relationships governing EEG signals. This sequential relationship being modelled in the temporal (EEG signals can be considered as a set of time series), spatial (EEG are recorded in several locations of the participant scalp) and frequential (EEG can be filtered in different frequency bands each of them being responsible for human behaviour).

In this context, due to the encouraging results of the novel techniques aiming to improve the analysis of sequential information: written sentences \cite{devlin_bert_2019} or speech segments \cite{chi_audio_2021}, it has been thought to merge this advanced to improve participant attention estimation from EEG. Our approach is based on the self-attention transformer encoder layers \cite{vaswani_attention_2017} allowing us to combine information from a non-neighbour element in a sequential signal which was not the case with conventional RNN as reported in \cite{merkx_comparing_2020}. Transformer based model can automatically process the sequential information from frequential bands, temporal windows or electrode location. In this work, feature matrices have been extracted from EEG data and represented in a 3D frame with three specific dimensions: temporal, spatial and frequential. This novel matrices representation is specially dedicated to our transformer architecture and aims to estimate attention state. The contributions of this work are the following: 1) creating transformer-inspired architecture suiting with EEG; 2) developing a three-stream network aiming to estimate attention state; 3) assessing the effect of frequential bands, temporal windows length and electrodes location; 4) finally, proposing a method that presents encouraging results exceeding the state-of-the-art approaches; 5) making extensive analysis of the feature extraction methods.

\section{Related Work}
\label{sec:rel}

During the last decade several research projects considering EEG signals have been completed. These last use ML algorithms for different estimation \cite{li_novel_2020,zou_eeg-based_2020,zhong_eeg-based_2020,lawhern_eegnet_2018,bashivan_learning_2016,cheah2020convolutional,ahangi2013multiple}. One of the specific subset of models consider the use EEG to retrieve the attentional or vigilance state of participants, i.e. focus vs. distract \cite{liu_eeg-based_2019,liu_inter-subject_2020,zou_eeg-based_2020}. 

Commonly, the first step consists of the preprocessing corresponding to band-pass filtering with or without ocular artefacts removing methodology. After, feature extraction often consists to frequential feature  extraction \cite{li_novel_2020,bashivan_learning_2016, shi_differential_2013,zheng_multimodal_2017}. However, some research projects proposed also an approach based on an automatic feature extraction methodology made with other DL models, i.e. deep-autoencoder \cite{wen_deep_2018}, or with larger DL architecture to automatically extract features for classification/regression \cite{lawhern_eegnet_2018}. Although this method is mainly employed in many other fields, it remains difficult to consider EEG processing without a handcrafted feature extraction step due to the signals' nature and the relatively small size of the public datasets. Most of the proposed approaches consider frequential based feature extraction methods to process EEG, however, other methods reflecting different signal properties can also be considered: signal's disorder with fractal dimension \cite{david_combined_2020}, temporal domain properties \cite{hjorth1970eeg, erdamar2012wavelet}.

From the computed feature arrays, it exists many representations based on methods originally dedicated to other tasks that have been adapted to EEG, e.g. image \cite{delvigne_attention_2020,bashivan_learning_2016} or graph \cite{zhong_eeg-based_2020}. The proposed representation in this paper will take into account the interdependence that exists among EEG signals in the spectral, temporal and spatial domains.

One of the main challenges this paper aims to tackle is the management of the sequential aspect of the considered inputs, i.e. considering the best ML-based approach to benefit from the relationship among input signals. In particular, this challenge is to find the best way to express the spatial (through electrodes), frequential (through frequential bands) and temporal (through the signal’s temporal evolution) relationship between EEG signals. Among the existing work, different methodologies have been proposed to solve these issues but often consider only the spatial information, i.e. how to organise the information to consider electrodes positions on the scalp:
\begin{itemize}
    \item Recurrent Neural Networks (incl. LSTM and GRU) that process spatial information in a unidirectional pathway. It is then necessary to consider one RNN for each direction (EEG spatial relationship being in two dimensions). Moreover, these models aim to estimate the recurrence in the sequential information, in the case of longer sequences the relationship between elements too far apart may not be taken into account \cite{merkx_comparing_2020}.
    \item Convolutional Neural Network can be used to model the spatial \cite{jia_sst-emotionnet_2020,yuan_hybrid_2019} by considering a 2D representation of EEG feature matrices \cite{jia_sst-emotionnet_2020}. An improved method aims to take into account the position of the electrodes by creating an image based on the interpolation of the location of the electrodes in the 3D frame \cite{bashivan_learning_2016,delvigne_attention_2020}. Another approach consists to extract temporal information from raw signals in the temporal and spatial domain by considering two-dimensions kernels \cite{lawhern_eegnet_2018}.
    \item Graph neural networks are a type of neural network that considers inputs as a graph. In the context of EEG, each electrode is considered as a node and the edges are proportional to the distances between them \cite{zhong_eeg-based_2020}.
\end{itemize}

It is important to note that the above-mentioned approaches are not necessarily implemented straightforwardly. Some works proposed a different approach consisting of concatenation or parallelized models. On the other hand, it exists methodologies considering a novel approach to improve the baseline results. For instance, by considering images-based EEG with CNN but with the help of self-attention mechanism on the spatial and temporal stream to increase the classification accuracy \cite{jia_sst-emotionnet_2020}.

Although several already presented approaches show high accuracy for EEG classification/regression in most of the cases they only consider an interpolated or uni-dimensional relationship between sequential information. For this reason, it has been thought to consider a novel approach for attention estimation by considering the sequential aspects in three directions: spatial, temporal and frequential.

\section{Proposed Method}
\label{sec:prop_meth}
In this paper, we proposed an innovative model aiming to estimate the attention state from EEG. The proposed approach to estimate attention is inspired by the transformer encoder from self-attention based model \cite{vaswani_attention_2017}. The motivations behind the use of this kind of model are justified by their ability to extract sequential information from different modalities \cite{vaswani_attention_2017,devlin_bert_2019, chi_audio_2021}. The proposed pipeline is separated into four steps: signals preprocessing, segmentation and representation; features extraction; modalities classification.

The preprocessing step follows the general recommendations for reproducible EEG research \cite{pernet_issues_2020}. The EEG dataset can be considered as a set of segmented signals $X^r$ $= [S_1 ^r,$ $S_2 ^r,$ $\dots , S_C ^r] \in \mathbb{R}^{C \times T}$ with $C$ and $T$ representing respectively the amount of electrodes on EEG recorder and the length of the signal. A bandpass filtering has been applied on each segment between $0.5$ and $50$ Hz. The lower band removes the continuous contribution and detrends the signals, the higher band removing electrical artefacts oscillating at $50$ Hz and a part of the muscular artefacts. An FIR filter with a Hanning window of 1-second has been considered for bandpass filtering. Another removing artefact methodology consisting of a manual removing signals by visual inspection and the use of the Automatic Artefact Removal (AAR) plug-in from EEGLab \cite{delorme_eeglab_2004} has been applied to remove the remaining ocular and muscular artefacts in $X_r$. This step is repeated for each trial and electrode, the preprocessed dataset can be reformulated as a matrix of dimension $[n_{trials} \times C \times T]$ with $n_{trials}$ being the number of trials during the total acquisition.

On the other hand, to compare signals corresponding to a high/low attention state, it is necessary to compute a label representing this feature. For this purpose, one physiological measurement correlated with the attention state have been considered: the reaction time during sustained-attention task, i.e. the time taken by a participant to react to a stimulus has been recorded. We consider the median for all the trials participants dependent (trials corresponding to the participant) and independent (all the trials). Then, a threshold is deduced for each participant by computing the mean between the median participant dependent and independent. Finally, in a trial corresponding to a physiological measurement above (resp. below), the threshold is considered as a low (resp. high) attention state. A binary class is then assigned to each trial.

As spectral information plays an important role in attention estimation \cite{delvigne_phydaa_2021, zheng_multimodal_2017}, it has been though to filter the signals into several frequency intervals, the latter may be physiologically pre-defined frequency bands (i.e. $\delta, \theta, \dots$ bands) or regular spectrum decomposition between $0$ and the cut-off frequency. Finally, the preprocessed EEG dataset is re-expressed by considering the band filtering as a set of signal $X_f^r $ $= [S_{1, f} ^r,$ $S_{2, f} ^r,$ $\dots , S_{C, f} ^r] \in \mathbb{R}^{F \times C \times T}$ with $S_{i, j} ^r$ being the EEG segment of $i$-th channel and $j$-th frequency band and $F$ being the amount of considered frequential bands. The filtering being made with same filter parameters.

After separating EEG into frequential contributions, EEG segments have been segmented into time windows. The goal of this segmentation is to capture the information from the signal's temporal evolution during task processing. Studies have shown that specific patterns occur in EEG during the sight of a stimulus \cite{lawhern_eegnet_2018}. The novel signal representation is $X_f^t$ $= [S_{1, f} ^t,$ $S_{2, f} ^t,$ $\dots ,$ $S_{C, f} ^t] \in \mathbb{R}^{F}$ $^{\times C \times T' \times \frac{T}{T'}}$ with $T'$ being the amount of temporal window, the length of the segment after the temporal segmentation is $\frac{T}{T'}$.

Then, from $X_f^t$, feature are extracted to express the signal in a shorter subspace. In the context of attention estimation, different feature extraction methods have already been considered each of them considering specific signals' aspect: Differential Entropy (DE) \cite{shi_differential_2013}; Fisher Information (FI) \cite{martin_fishers_1999}; Hjorth parameters \cite{hjorth1970eeg}; Petrosian Fractal Dimension \cite{mardi2011eeg}; Teager Energy \cite{hamila1999teager}.

Finally, from the feature matrice $F \in \mathbb{R}^{F}$ $^{\times C }$ $^{\times T'}$ $^{\times n_{feat}}$, it is possible to consider its representation as a sequence in three dimensions: frequential, temporal and spatial. The frequential direction takes into account the signal evolution among the considered frequential bands. The spatial direction represents the electrodes based relationship between features information and depends on the order in which the electrodes are sorted. Finally, the temporal dimension expresses the time-based evolution of each feature vector. Practically, this representation can be expressed considering each of this dimension, after transposing and merging axes, the resulting representation can be expressed as: $ EEG|_{frequency} \in \mathbb{R}^{F \times n_{feat-freq}}$, $EEG|_{temporal}\in \mathbb{R}^{T' \times n_{feat-temp}}$, $EEG|_{spatial}\in \mathbb{R}^{C \times n_{feat-spat}}$.

The feature dimension for each of these three representations is deduced from the reshaped dimension of the feature matrice $F$. The interest of this representation is that it permits to have a sequential representation of information considering each stream (i.e. frequential, temporal and spatial) separately. Moreover, this representation allows not to lose information or limit biases, unlike for instance the image-based representation that considers an interpolation of a feature map.

\begin{figure}[!t]
    \centering
    \includegraphics[width=2.85in]{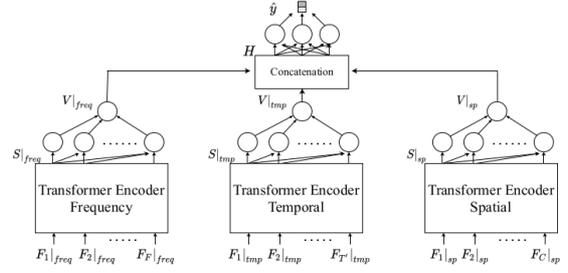}
    \caption{Overview of the proposed architecture for attention estimation. The three representations of the EEG features arrays are passed to the correspond transformer encoder stream. Then the outputs are decoded and concatenated to create an hidden vectors passed to a fully-connected networks to estimate the attention state.}
    \label{fig:ModelOverview}
\end{figure}

It exists several DL based algorithms to estimate modalities from this particular representation of information \cite{lotte_review_2018}. In this paper, an adapted version of the encoder layers from the transformer architecture \cite{vaswani_attention_2017} is proposed. This architecture is composed of different blocs each of them being responsible for a specific aspect. If we consider the input feature matrix $F$ representing the sequential information in one of the three dimensions as explained in the previous subsection, the estimated class $\hat{y}$ is computed after the following steps:
\begin{itemize}
    \item Embedding aiming to have a continuous representation of the feature in a vector of lower dimension;
    \item Positional encoding allowing to add information about the element position in the sequence given that self-attention mechanism providing not information about recurrence in the sequence (unless RNN).
    \item Transformer encoder applying attention mechanism on all the previous elements composing the sequence resulting in a hidden representation of the input embedded vector.
    \item Adapted transformer decoder consisting of a feed-forward network (FFN) applied on the hidden vectors resulting from the encoder. A second FFN merges these vectors to provide a single representation for each sub-model corresponding to signal representation as seen in Fig \ref{fig:ModelOverview}.
    \item Finally, the resulting three vectors are concatenated and passed through a FFN aiming to estimate attention state. 
\end{itemize}

The proposed architecture aims to estimate attention state from feature arrays computed from EEG. The goal of the training phase is to find the correct value for each trainable parameter to make the correct estimation. The considered loss consists of a categorical cross-entropy.

\section{Experiments}
\label{sec:exp}
In this section, we describe the considered datasets and models in this paper, as well as the settings and parameters for attention estimation from EEG.

\subsection{Datasets}

In our experiments, two different datasets of EEG signals have been considered: PhyDAA \cite{delvigne_phydaa_2021} and Driving EEG \cite{cao_multi-channel_2019}. Their goal is to proposed segmented signals corresponding to a specific attention state. The methodology employed to assess attention state is based on the registration of the reaction time to specific stimuli. These stimuli can be represented by a balloon appearing inside of virtual reality (VR) environments \cite{delvigne_phydaa_2021} or by steering angle modification of a car during a driving task \cite{cao_multi-channel_2019}. PhyDAA dataset proposed an experiment during which participants are asked to react as fast as possible to a specific stimulus. The reaction is measured with the direction of the eyes, it corresponds to the time elapsed to direct the sight toward the stimuli. 32 participants took part in the 15 minutes length experiments. Driving EEG dataset consists of an attention assessment during driving task \cite{cao_multi-channel_2019}. This dataset proposes a task in VR environment representing a car driving task, during which it is asked to react as fast as possible to perturbators corresponding to the deviation of the car trajectory. The time taken to correct the steering angle is jointly measured. 27 participants took part in the 90 minutes experiment. 

The 32 electrodes have been placed following the 10/20 disposition for both datasets and registered at a sampling frequency of 500 Hz. The steps already presented in the third section have been applied to extract the feature and split the feature arrays in each of the three directions (frequency, temporal and spatial). To investigate the effect of frequential bands and temporal windows, it has been decided to divide the samples into $[1, 4, 10, 20]$ temporal windows and have been filtered in $[1, 5, 20, 50]$ frequential bands\footnote{The computed value for the lengths of both windows have been chosen after a preliminary study.}.

\begin{table}[t]
    \centering
    \begin{tabular}{c c c}
    \hline
    \hline
    \rule{0pt}{4ex}  
    Approach & Driving EEG \cite{cao_multi-channel_2019} & PhyDAA \cite{delvigne_phydaa_2021}  \\
    & ACC/STD [\%] & ACC/STD [\%] \\
    \hline
    TCA + LR \cite{liu_eeg-based_2019} & 72.70/9.42 & - \\
    MIDA \cite{liu_inter-subject_2020} & 73.01/9.17 & - \\
    Graph Network \cite{delvigne_phydaa_2021} & - & 72.41/5.51 \\
    \hline 
    SVM$^*$ & 68.09/9.55 & 64.61/9.22 \\
    RF$^*$ & 67.81/10.17 & 61.55/9.79 \\
    RNN$^*$ & 72.12/8.27 &  70.86/9.82 \\
    ResNet$^*$ & 62.07/6.20 & 66.82/5.21 \\
    \textbf{Transformer}$^*$ & \textbf{74.41/9.27} & \textbf{77.24/6.11} \\
    \hline
    \hline
    \end{tabular}
    \caption{Classification performance of the different methods considering participant independent cross-validation, i.e. with leave one subject out cross-validation accuracy. * denotes the results obtained from our models experiments. }
    \label{tab:resMod_Ind}
\end{table}

\subsection{Settings}

The model evaluates with subject dependent and independent has been configured with the same parameters. The chosen dimensions were respectively equal to $F = [1, 5, 20, 50]$, $T' = [1, 4, 10, 20]$ and $C = 32$ for each of the three dimensions. The transformer encoder part of the architecture is composed of two transformer encoder layers each of them composed of four heads in the multi-attention model part \cite{vaswani_attention_2017}. The chosen dimension for the embedded representation and the dimension of the self-attention matrices is equal to $64$ and $128$. The training has been made considering a stochastic gradient descent (SGD) optimizer with a scheduled learning rate beginning at $1e-2$ with $\gamma=0.99$. The batch size and number of epochs are respectively $32$ and $250$. The model has been implemented using Pytorch library and the training has been made on one Nvidia Titan RTX GPU. For sake of reproducibility, the model's implementation and the codes used for the preprocessing are freely available on github\footnote{ \url{https://github.com/VDelv/Spatio-Temporal-EEG-Analysis}}.

\section{Results}
\label{sec:res}

In this section, we discuss the results achieved to retrieve the attention state. A comparison with other methodology, a study of the different chosen parameters, the activation maps resulting and an ablation study has been performed.

\subsection{Comparison of deep learning models}

\begin{table}[b]
    \centering
    \begin{tabular}{c c c}
    \hline
    \hline
    \rule{0pt}{4ex}  
    Approach & Driving EEG \cite{cao_multi-channel_2019} & PhyDAA \cite{delvigne_phydaa_2021}  \\
    & ACC/STD [\%] & ACC/STD [\%] \\
    \hline
    MLP \cite{zou_eeg-based_2020} & 81.32/6.02 & - \\
    Graph Network \cite{delvigne_phydaa_2021} & -  & 77.34/10.24 \\
    \hline 
    SVM$^*$ & 76.07/9.65 & 70.82/13.25 \\
    RF$^*$ & 75.60/8.76 & 75.63/12.89 \\
    RNN$^*$ & 80.03/8.09 & 79.64/10.55\\
    ResNet$^*$ & 75.96/8.98 & 70.39/6.91 \\
    \textbf{Transformer}$^*$ & \textbf{83.31/6.71} & \textbf{85.04/7.56} \\
    \hline
    \hline
    \end{tabular}
    \caption{Classification performance of the different methods considering participant dependent cross-validation. * denotes the results obtained from our models experiments.}
    \label{tab:resMod_Dep}
\end{table}

To evaluate the proposed methodology, the architecture has been trained and validated with two different datasets. Two training methodologies aiming to assess the model faculty to generalise have been considered in this paper: 1) Subject-Independent classification, where the model is trained with all the participant signals except one that is used for the validation and the step is repeated for all subjects and a mean cross-validation accuracy and its standard deviation is computed. The benefit of this method is to measure the model ability to generalise its knowledge to never met participants; 2) Subject-dependent classification where the model is trained and validated with the same participant following a regular 5-fold cross-validation, the process is repeated for each participant and the mean and standard deviation of cross-validation accuracy are computed. The advantages of this method are that it gives a good insight into the model ability to make estimations with fewer signals. 

It was also thought to consider the comparison with the different methodology aiming to estimate attention from feature matrices constructed from EEG signals. Among the existing ML models, four have been considered: 
\begin{itemize}
    \item Traditional ML models: Random Forest (RF) and Support Vector Machine (SVM) based classifier to define a baseline result for attention estimation.
    \item RNN based approach consists of the transformer-based approach represented in Figure \ref{fig:ModelOverview} where the transformer encoder layers are replaced by RNN for each stream.
    \item CNN approach based on an image-based representation of the EEG feature map. Then the images are passed through a resnet architecture \cite{he_deep_2016}.
\end{itemize}

As seen in Tables \ref{tab:resMod_Ind} and \ref{tab:resMod_Dep}, results acquired from the transformer-based approach present the highest accuracy compared to other baseline approaches for both datasets that demonstrate the proposed framework's ability to estimate attention from EEG. More, it shows the efficiency of self-attention based models architecture to process sequential signals. 

Furthermore, results from previous works have also been compared to evaluate the proposed methodology. For the first dataset, the specificity of the previous approach is based on a transfer learning approach to increase cross-subject accuracy. These last are based on Transfer Component Analysis (TCA) or Maximum Independence Domain Adaptation (MIDA) with traditional ML architecture that may cause an accuracy decay compared to the more complex methods. As seen in Tables \ref{tab:resMod_Ind} and \ref{tab:resMod_Dep} the results from our experiments from traditional ML approaches, i.e. SVM and RF, are lower compared to other DL methods. It makes us think that considering a more complex training methodology, including transfer learning, may lead to an increase in the transformer accuracy, although its accuracy is outperforming the state-of-the-art models. 

For the second dataset, the best results from the related works are based on Graph Convolution Network (GCN). Its architecture is composed of graph convolution and a pooling operation aiming to keep only the most discriminant electrodes. Unlike the transformer, GCN only considers the spatial stream to estimate attention from EEG. This may explain the lower results. 

\subsection{Feature parametetrs analysis}

\begin{figure}
    \centering
    \includegraphics[width=2.75in]{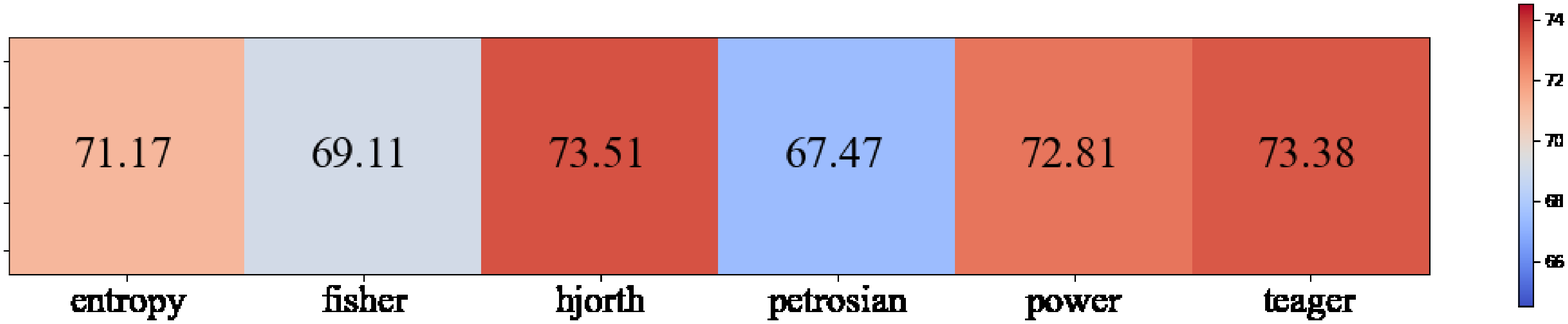}
    \hspace{2 cm}
    \centering
    \includegraphics[width=2.25in]{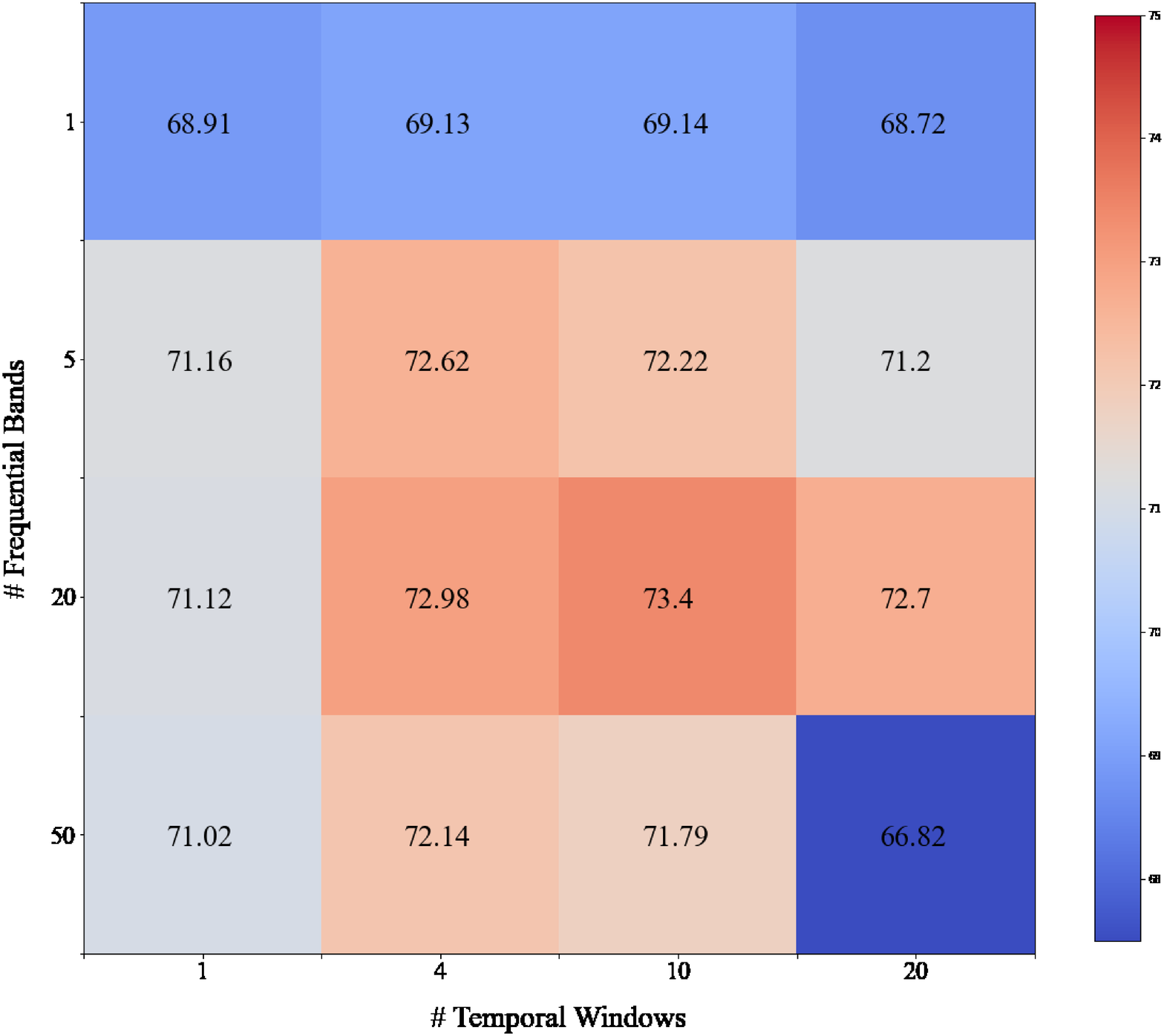}
    \caption{Mean cross-validation accuracy in function of the feature extraction methods (above); the  amount of temporal windows and frequential bands (below).}
    \label{fig:feat_analysis}
    
\end{figure}

As mentioned in Section \ref{sec:prop_meth}, different feature extraction methods and segmentation parameters have been considered. In this section, we present the corresponding cross-validation accuracy for each combination. In Figure \ref{fig:feat_analysis}, the feature extraction methods present cross-validation accuracy around similar range of value $\approx 70 \%$. As shown, Hjorth, TE, PSD and DE present the best results. Moreover, the two best feature extraction methods: Hjorth parameters and TE based operator, consider both the signals' derivative that corroborates the fact that the derivative play an important role in the attention estimation from EEG. 

As seen, the amount of both time windows and frequency bands play an important role in attention estimation. As seen in Figure \ref{fig:feat_analysis}, for both number of temporal windows and frequential bands a too small number of time windows/frequency bands leads to a decrease of accuracy. This decay can be caused by the difficulty of representing the evolution of the brain activity during the stimuli apparition or among the spectrum. More, a too large number can lead to a decrease in accuracy due to overfitting issues.

The medium values present the higher results temporal and spectral parameters. More precisely, better results are proposed for regularly cut bands (i.e. with $20$ for \#Frequency Bands) compared to pre-defined bands (i.e. with $5$ for \#Frequency Bands). This can be explained by the fact that some populations do not present the same bands limits the pre-defined \cite{bazanova_individual_2010, arns_decade_2013}.

\subsection{Ablation studies}

\begin{figure}[t]
    \centering

    \begin{minipage}[b]{1.4in}
        \includegraphics[width=\textwidth]{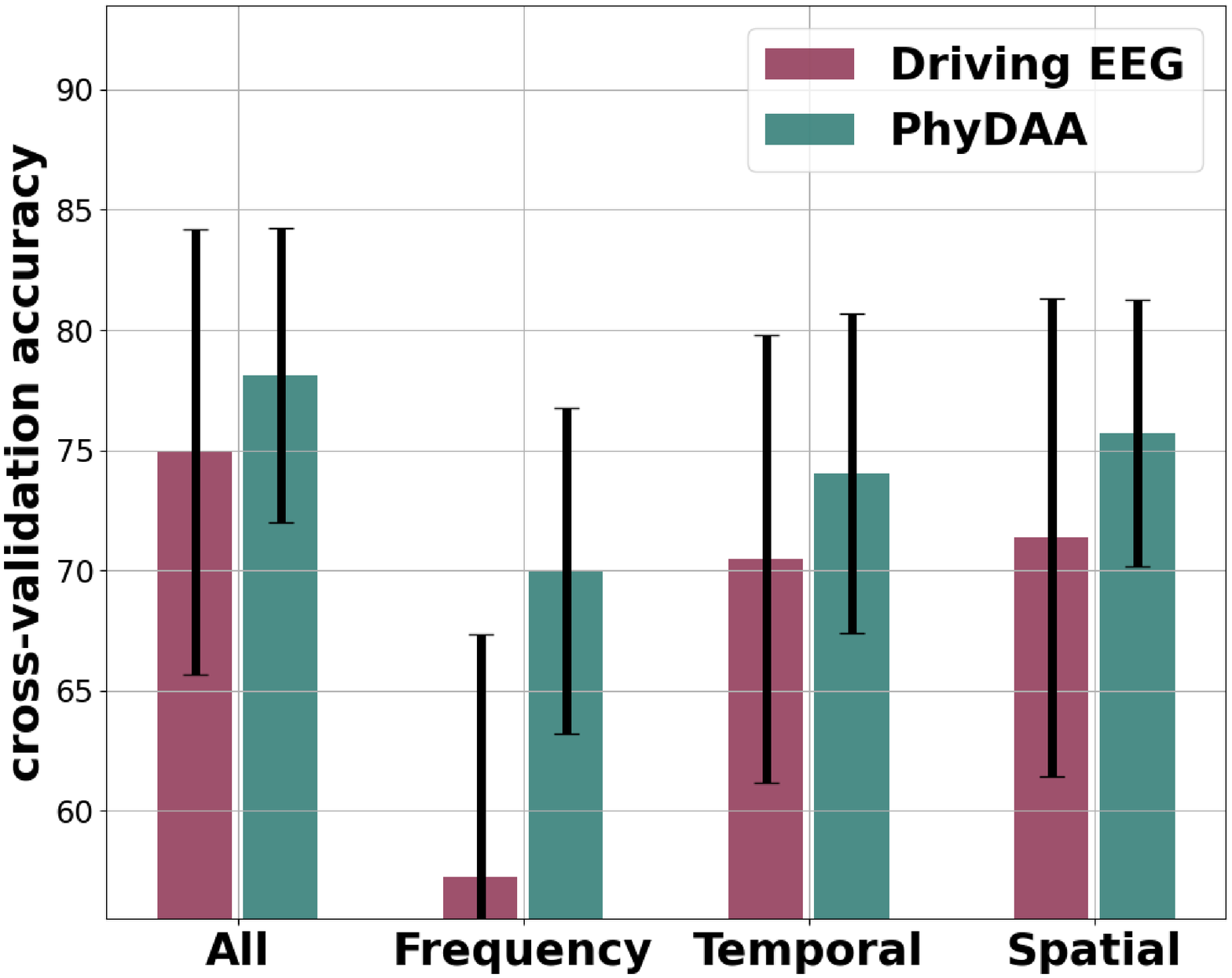}
        \caption*{(a) Subject Independent.}
    \end{minipage}
    \hfill
    \begin{minipage}[b]{1.4in}
        \includegraphics[width=\textwidth]{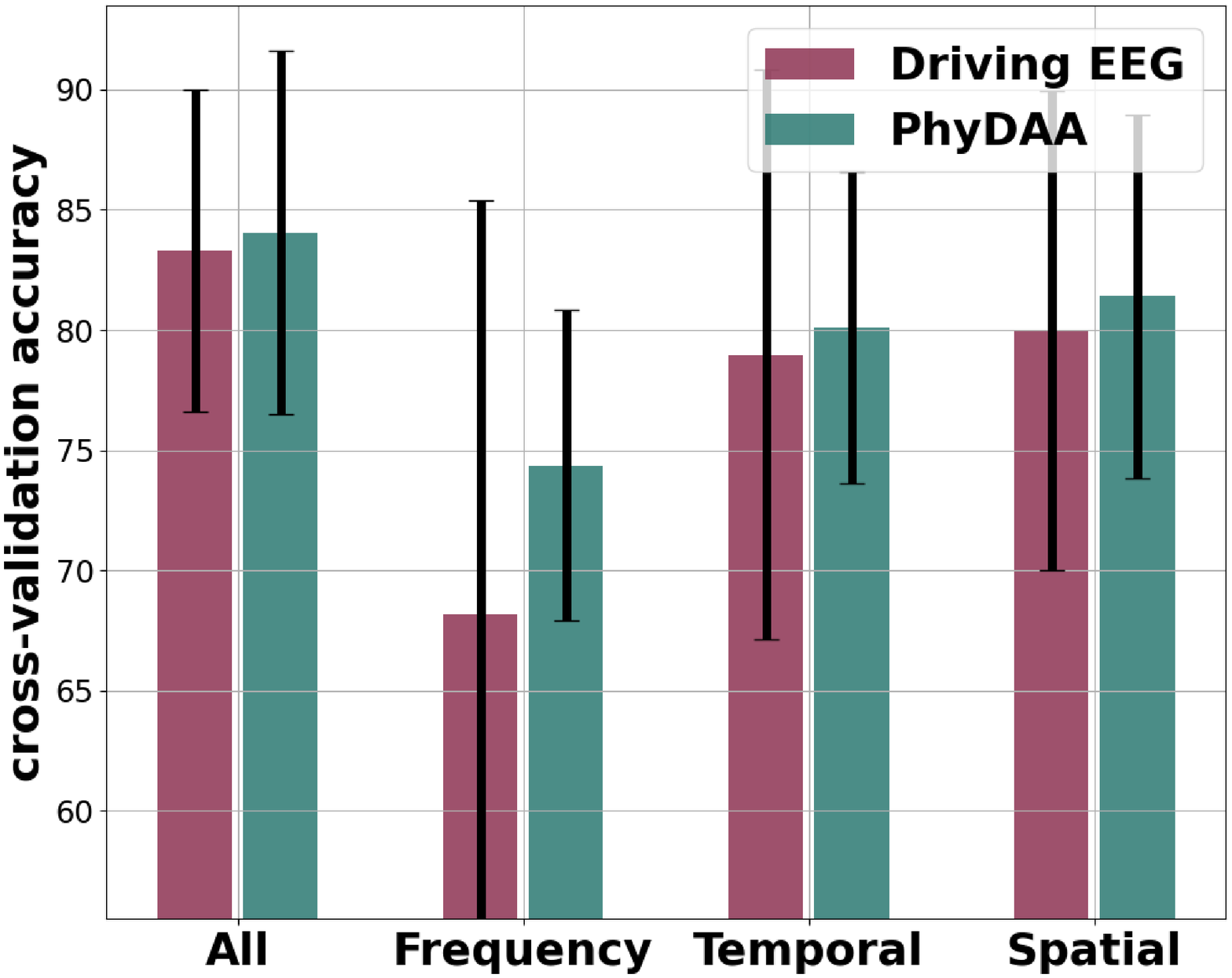}
        \caption*{(b) Subject Dependent.}
    \end{minipage}
    
    \caption{Ablation study for the transformer models. The left (resp. right) figure correspond to the subject independent (resp. dependent) mean cross-validation. The bars colour correspond to the considered datasets with each bar corresponding to a stream.}
    \label{fig:stream_comp}
\end{figure}

In addition to the comparison based on the considered architecture or signal parameters, it has been thought to consider a comparison aiming to investigate the contribution of each stream. For this purpose four different architectures have been considered: 1). the original approach as described in section \ref{sec:prop_meth} considering the concatenation of the three streams; 2). frequential; 3) temporal; 4) spatial based transformer stream standalone. The experimental results from these four different approaches are listed in Figure \ref{fig:stream_comp}. As seen, in both of the cases and datasets, the best results were noted for the approach considering the information from the three directions. This observation corroborates the fact that considering all the available information is the best approach to estimate attention from EEG.

The best results for a single direction based approach were acquired by considering only the spatial stream. The temporal based approach presented slightly lower results. These findings may be explained by the fact that the spatial, i.e. electrodes-based, and temporal information have played an important role in the attention estimation. The activation areas deduced from EEG are considered as a good biomarker for behaviour/movements estimation \cite{edelman2015eeg}, which may explain the high accuracy provided by spatial information. The importance of the temporal information and the resulting scores are explained by the nature of the signals composing the datasets. In both of them, each segment can be considered as Event-related potentials (ERP), i.e. the brain response resulting from a stimulus \cite{ogryzko_transcriptional_1996} and present a specific pattern. In the context of this experiment, stimuli apparition is fixed at $t=1$ second, however, ERP may appear at different instants depending on the attention state. The frequential based approach acquired the lowest accuracy for both dataset and training methods as mentioned in Figure \ref{fig:stream_comp}. The consideration of this stream is motivated by the fact that previous works mentioned the importance of frequential information to retrieve attention state, and especially the middle bands \cite{bioulac_personalized_2019}. Poorer results can be expressed by the redundancy of the spectral information already extracted during feature extraction.

\subsection{Activity maps analysis}

To investigate the contribution of each stream, it has been thought to analyze the activity maps generated by the transformer-encoder. For this purpose we consider the ${L}_2$ norm of each output sequence from transformer encoder for each stream, i.e. $S_i|_{freq}, S_i|_{temp}, S_i|_{spatial}$ as shown in Figure \ref{fig:ModelOverview}. This process has been made during the training of the three-stream model training and the weight have been frozen. After considering the norm of every sequence, they have been normalized to study their contribution. 

\begin{figure}
    \centering
    \includegraphics[width=2.25in]{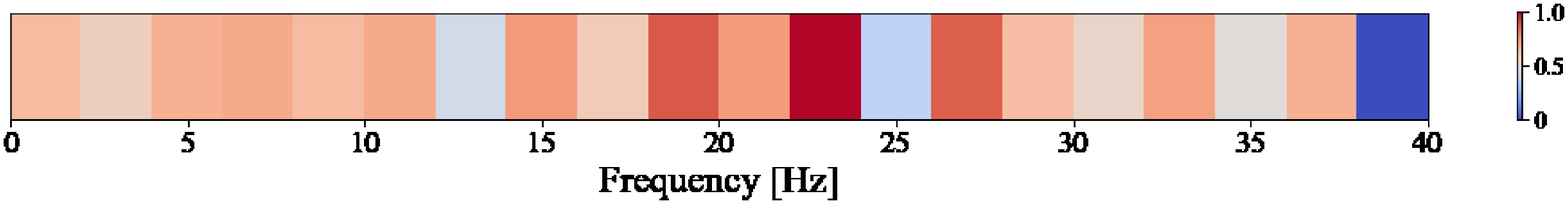}
    \hspace{2cm}
    \centering
    \includegraphics[width=2.25in]{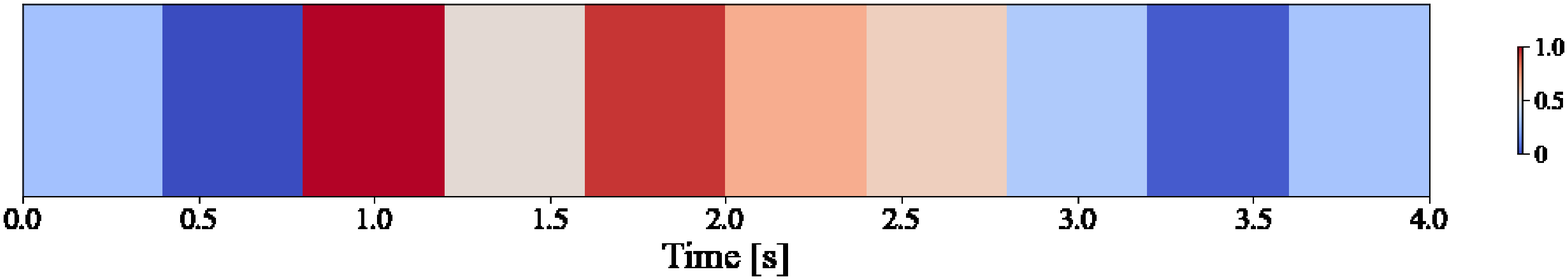}
    \hspace{2cm}
    \centering
    \includegraphics[width=2in]{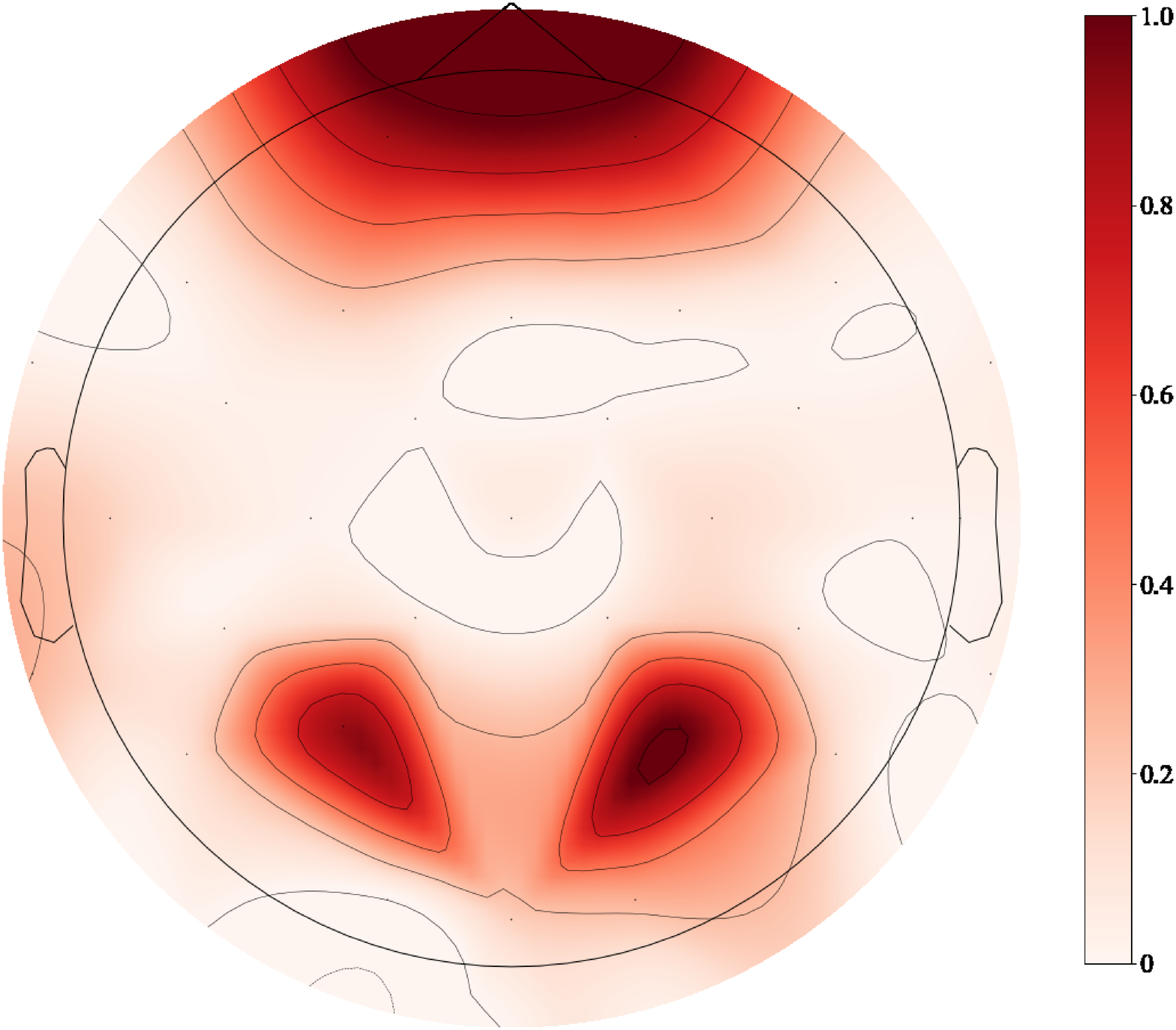}
    \caption{Activation map in function of attention state classification of the hidden representation of the corresponding representation: Top - Spectral; Middle - Temporal; Bottom - Spatial. These maps have been normalized to reflect the contribution of each temporal windows.}
    \label{fig:activation_map}
\end{figure}

In Figure \ref{fig:activation_map}, the activity map resulting for the spectral stream is displayed. As seen, three frequential areas emerge from the activity spectrum. First, the sparsity of the high-frequency contribution (i.e. $>25$ Hz) and their irregularity among the area makes the author think that they are caused by the remaining artefacts from muscular activity that have not been removed from the pre-processing step but can be correlated with attention state as it has already been proven \cite{blume_nirs-based_2017}. The two other spectral bands with high activity superposed with the physiologically defined frequency bands $\theta, \alpha$ and $\beta$ that are often used for attention estimation \cite{bazanova_individual_2010,liu_eeg-based_2019,liu_inter-subject_2020}. More, the behaviour related to the task proceeds by the participant is related to the theoretically defined behaviour by this frequency bands \cite{collura_chapter_2009} that corroborates these insights.

As seen in Figure \ref{fig:activation_map}, the most salient instant during the trial is between $1$ and $2.75$ seconds with higher importance between $1$ and $2$ second that corresponds to the instant directly following the stimulus apparition. This period seems therefore important to distinguish high/low attention segments.

At the bottom of Figure \ref{fig:activation_map}, the most salient EEG-based region are displayed. First, a casi symmetry is observed between the two hemispheres that make it possible to reject an electrode misplacement or mis-conduction due to the specific registration conditions. More significantly, two different regions stand out from the spatial activity: frontal and parietal regions from electrodes placements. In addition to being one of the most salient electrodes regions for attention estimation, it has been shown that the parietal region is also responsible for attention mechanism \cite{behrmann2004parietal}.

\section{Conclusion}
\label{sec:concl}

In this work, we present a framework aiming to estimate the attention state from EEG signals during specific tasks. We propose a novel approach to handle these signals based on a three-fold information representation based on the frequential, temporal and spatial features. Moreover, a novel Transformer inspired architecture for EEG processing has been presented. This last allows extracting the sequential information from the EEG feature maps in each of the three dimensions mentioned above. To validate this new method, the framework has been trained and tested on public datasets assessing the attention state. The results are encouraging and outperform the state of the art approaches. The proposed models can be useful for different applications such as attention assessment for subjects with ADHD to detect and help to reduce their symptoms; another application could be a vigilance estimator during driving to alert the driver in case of drowsiness. In further works, we want to explore other EEG datasets to investigate the feasibility of a large framework that can be applied to various fields, more the application of the model in a real-life application will be considered. Over the next years, we think that the use of EEG and ML models will be helpful to help in diagnosis and treatment, and to prevent accidents. 

\bibliographystyle{IEEEtran}
\bibliography{IEEEfull}

% that's all folks
\end{document}